\documentclass[12pt]{article}
\usepackage{latexsym,amsmath,amsfonts,natbib,bm}
\usepackage{bbm}
\usepackage{amssymb,amsthm,dsfont}
\usepackage{graphicx}
\usepackage{mathtools}
\usepackage{setspace}
\usepackage{appendix}
\usepackage{comment,verbatim}
\usepackage{microtype}
\usepackage{xcolor}
\usepackage{multirow}
\usepackage{subcaption}
\usepackage{float}
\RequirePackage{hyperref}

\DisableLigatures[f]{encoding = *, family = * }
\numberwithin{equation}{section}

\def\balpha{\bm{\alpha}}
\def\bbeta{\bm{\beta}}
\def\btheta{\bm{\theta}}

\def\bX{\bm{X}}
\def\bH{\bm{H}}

\makeatletter

\def\@seccntformat#1{\@ifundefined{#1@cntformat}%
   {\csname the#1\endcsname\quad}  % default
   {\csname #1@cntformat\endcsname}% enable individual control
}
\let\oldappendix\appendix %% save current definition of \appendix
\renewcommand\appendix{%
    \oldappendix
    \newcommand{\section@cntformat}{\appendixname~\thesection\quad}
}
\makeatother

\title{\bf Modern strategies for time series regression}
\author{Stephanie Clark\thanks{School of Mathematical and Physical Sciences, University of Technology Sydney, Australia.},
Rob J Hyndman\thanks{Department of Econometrics and Business Statistics, Monash University,  Australia.},
Dan Pagendam\thanks{Data61, Commonwealth Scientific and Industrial Research Organization, Australia},
Louise M Ryan\thanks{School of Mathematical and Physical Sciences, University of Technology Sydney, Australia.}\footnote{Communicating author: \href{mailto:louise.m.ryan@uts.edu.au}{\tt louise.m.ryan@uts.edu.au}}}
\date{}
\begin{document}
\maketitle

\begin{abstract}
This paper discusses several  modern approaches to regression analysis involving 
time series data where some of the predictor variables are also indexed by time.  We discuss classical statistical approaches as well as methods that have been proposed recently in the machine learning literature.   The approaches are compared and contrasted, and it will be seen that there are advantages and disadvantages to most currently available approaches.  There is ample room for methodological developments in this area.  The work
is  motivated by an application involving the prediction of water levels as a function of rainfall and other climate variables
in an  aquifer in eastern Australia. 

\end{abstract}
\begin{flushleft}

\end{flushleft}
\doublespacing

\section{Introduction}
Statistical methods for the analysis and forecasting of time series data have a long history \citep{tsay2000time}. The well-accepted Box-Jenkins analysis and forecasting methods have been applied in a wide range of applications, from finance to medicine, and the classic book that laid out the theory is now in its fourth edition with over 55,000 citations \citep{box2015time}. 

In this paper, we focus on the specialized area of time series regression where the goal is to predict one time series with the help of covariates that include elements which also have a time series nature.   Some authors refer to this as dynamic regression \citep{HyndmanBook}, others use the term {\tt regARIMA} \citep{Gomez1994-hx, Maravall2016-hy}. \cite{pankratz} provides an excellent overview.   

Over the past several decades, alternatives to the classical statistical approach to time series modelling have emerged in the machine learning literature. See \cite{Goodfellow-et-al-2016} for a readable introduction to deep learning approaches, including their generalizations to the time domain. While several papers have compared the approaches \citep{BontempiEtal,MakridakisEtal}, the treatment has been fairly limited and for the most part has not considered the context of interest here in which predictor variables are also included.  

Our work is motivated by a project to develop prediction models for water levels in underground aquifers so as to better understand how these levels are influenced by rain and other climate features. Efforts like these are  becoming increasingly important in the global response to
the challenge of managing water resources in a sustainable way,
especially in the face of increasing populations and changing climate \citep{bogardi2012water, tardieu2018global}. Water resource management is a particular challenge in Australia, which is known to be the dryest continent on earth, aside from Antarctica, (\url{https://www.ga.gov.au/home}) and is vulnerable to large scale oceanic patterns that can lead to extended periods of drought and high temperatures. Since many parts of Australia rely on aquifers for drinking water and irrigation,
reliable monitoring methodologies are needed  
so that government agencies can plan in a sustainable way for the future.

The traditional approach to groundwater modeling involves the use of complex deterministic approaches based on physics, fluid mechanics and soil mechanics  \citep{langevin2017documentation}. These models require specific knowledge of the local underground characteristics in the vicinity of the aquifers. While in theory the required data can be collected, doing so is expensive and time consuming. For this reason, there has been increasing interest in recent years in the development of more data-driven approaches that utilize statistical and machine learning strategies \citep{ BakkerSchaars,kratzert2018rainfall,zhang2018developing}. The overall purpose of this paper is to discuss, compare and evaluate some of these strategies.

Later in the paper, we will present a case study involving the development of prediction models for water levels measured in a monitoring bore in the Richmond River basin in the northeastern part of the Australian state of New South Wales (NSW). Although data are available from multiple monitoring bores in the Richmond River basin, we will focus for much of the paper on strategies for modeling the time series corresponding to data collected from a single bore. 

The paper is organized as follows.  After introducing some notation at the beginning of Section \ref{methods}, we briefly review some classic statistical approaches to time series regression analysis. We then turn to a brief overview of some machine learning approaches to prediction modeling based on covariates. There has been increasing interest in the use of neural networks for this task, especially in the context of large, complex datasets with many potential predictors. We outline the use of neural networks in standard regression, drawing extensively from Chapter 11 of \cite{HTF} which provides an exposition of the relationship between neural network models and more traditional statistical approaches. Those authors suggest that neural network models can be thought of as generalizations of projection pursuit models that were introduced in the  1980s \citep{FriedmanStuetzle}. We discuss some of the various extensions of neural networks that have been developed to accommodate time-indexed data.  Broadly speaking, these fall into the class of models called recurrent neural networks (RNNs).  
In Section \ref{simulations}, we present a simulation study based on two different models for data generation that draw on 
hydrological theory.  The first one uses the concept of transform functions \citep{montgomery1980modeling} to produce a fairly simple relationship between water level and rainfall. We will see that the resulting models are  relatively linear and there is little difference among the performance of the various modelling approaches. The second simulation draws on a more complex, physically-based approach for data generation using the catchment water balance model developed by \cite{perrin2003improvement}. This approach incorporates more non-linearity and  autocorrelation into the relationship between water level and the predictors of rainfall and evapotranspiration. We will see that with this simulation the modelling becomes more complicated and differences amongst the results of the various approaches emerge. Using this second more complex simulation, we explore the impact of various modelling choices on  prediction accuracy.   Section \ref{application} presents a real-world analysis of the effect of rainfall and  evapotranspiration on water levels in a monitoring bore in eastern Australia. Section \ref{discussion} concludes with a summary of our findings, some recommendations in practice, as well as suggested avenues for further research.

\section{Notation and Methods}
\label{methods} 
We begin by defining some notation.   Let $Y_t$ (with $t=1,...T$) represent the outcome of interest 
 at time $t$ and let $\mathbf{X}_t$ be a $p \times 1$ vector of the relevant predictors available 
 at time $t$.  
 Some of the elements in $\mathbf{X}_t$ may be fixed in time, whereas others will vary. 
For our bore water application, $Y_t$ will 
represent the water level measured in the bore of interest at time $t$, while $\mathbf{X}_t$ will represent rain and other relevant climate variables that will change over time.  While it is possible for $Y_t$ to be a vector of outcomes measured at time $t$ (in our application, for example, this
might represent the case where  multiple bore water levels were being modelled simultaneously), our discussion will focus on a single, univariate time series.  

In many practical settings,  relevant predictors at any given time $t$ will include things measured concurrently in time, as well as over recent history. In the context of air pollution modelling, for example, it is well known that daily hospitalisations for respiratory and cardiac conditions depend on air pollution levels over the past several days, not simply pollution levels on the day of admission \citep{Lalletal}.  
In our application, bore water levels on a particular day $t$ 
will be impacted by a cumulative effect of rain over preceding weeks or months, not just rainfall on day $t$. 
Indeed,  the determination and characterization of this relationship is the central element of the challenge we face here and we will be discussing it in much more depth presently.  For now, it is sufficient to indicate that $\mathbf{X}_t$ will include measurements on all the relevant predictors of interest at time $t$ and that this may well
include measurements taken not only on day $t$, but also at timepoints prior to $t$ as well.  
While the models that will be discussed presently in this section can implicitly capture some of these lag effects, we will see in our simulations, as well as in our application, that it will generally be much more effective to incorporate the lagged predictors directly into $\mathbf{X}_t$.  

\subsection{Classical time series modelling}
\label{classical} 

A full review of the massive literature on time series modelling is well outside the scope of this paper.  We focus instead on a particular specialised aspect where the outcomes of interest, $Y_t$ can be expressed as a function of the predictors, $\mathbf{X}_t$, plus an error term, $\eta_t$, that allows for autocorrelation: 
\begin{equation}
 Y_t = \bm{\beta} \mathbf{X}_t + \eta_t.
 \label{dynamicRegression}
 \end{equation}
 
\cite{HyndmanBook} refer to this as  
dynamic regression modelling  and \cite{pankratz} provides extensive discussion on this.

There are several {\tt R} packages available to fit these kinds of models, but we will be using the 
{\tt ARIMA} function from the forecasting package {\tt fable} (see \url{https://fable.tidyverts.org/}). This package allows  $\eta_t$  to have an ARIMA($p,d,q$) structure: 
\begin{equation}
 \left(1-\sum_{i=1}^p \phi_iB^i\right) (1-B)^d \eta_t = \left(1+\sum_{i=1}^q \theta_iB^i\right) \epsilon_t, 
\label{ARIMA}
\end{equation}
where $\epsilon_t$ is white noise and $B$ is the back operator such that
$B\eta_t = \eta_{t-1}$, $B^2\eta_t = \eta_{t-2}$, etc. Many familiar examples are special cases.  For example, a single-lag auto-regressive, or AR(1) model,  corresponds to ARIMA(1,0,0): $\eta_t = \phi \eta_{t-1} +  \epsilon_t$. A two-lag moving 
average model, or MA(2), 
corresponds to ARIMA(0,0,2): $\eta_t =  \epsilon_t + \theta_1 \epsilon_{t-1}  + \theta_2 \epsilon_{t-2}$.  
The {\tt fable} package allows the differencing parameter, $d$, to take the values 0, 1 or 2.  The special case ARIMA(0,1,0) corresponds to assuming that differences between the errors at two successive timepoints are white noise (this assumption is often used in modelling financial data). 
The  {\tt fable} package allows the user to either specify the values of $p, d$ and $q$ or the package can automatically select the values that yield the best fit in terms of AIC values.  

An alternative to the {\tt fable} package, the {\tt arfima} package, is slightly more general in that it allows arbitrary (non-integer) values of $d$. However, it does not have some of the other appealing aspects of the  {\tt fable} package, particularly in relation to forecasting.

In principle, the dynamic regression model (\ref{dynamicRegression}) 
can be generalized to allow a non-linear function of $\mathbf{X}_t$, with the linear term
$\bm{\beta} \mathbf{X}_t$ replaced by a smooth function 
$g(\mathbf{X}_t)$.  The {\tt mgcv} package in {\tt R} allows
the specification of an AR(1) error structure \citep{wood}, however to the best of our knowledge reliable software are not available for fitting such GAM-type models, whilst allowing for more general correlated error structures. One option would be to use a package such as {\tt fable}, but 
with an expanded predictor space that includes, for example, additional predictors created via spline basis functions.  However, space considerations preclude further exploration of this approach in this paper. 

As discussed extensively by 
\cite{HyndmanBook}, once a model has been fitted there are a number of different options for how one can predict future values of the time series for timepoints beyond the input range (that is, predicting $y_t$ when $t>T$).   These range from naive methods, such as last value carried forward,
to sophisticated methods that exploit the assumed ARIMA stucture of the error component of the model and appropriately incorporate uncertainty (see section 9.8 of 
\cite{HyndmanBook}). 

In practice, the most challenging aspect of dynamic regression modelling for time series will be properly specifying the predictor space.  As discussed earlier, this will be important in the context of our application where bore water levels at any given time
are likely to reflect a fairly complex composite of rainfall and other climatic effects over the preceding period of time, possibly months. 
One simple option is to ensure that the vector of predictors at time $t$ includes all the appropriate lagged variables, for example rain on day $t-1$, rain on day $t-2$ etc.  We will be exploring this and related approaches through our simulations, as well as in the application section of the paper. Such an approach has the advantage of being fairly straightforward, but it does result in models that are quite highly parameterized and co-linearity can become an issue. Also, including lag terms in this way
does not address the issue of possible non-linear effects and interactions. To incorporate these, one would have to explicitly construct the appropriate elements of the design matrix. 

Considering the approach of including lag terms explicitly in the model specification brings up another 
interesting aspect of dynamic regression modelling.  In particular, by creating some additional lagged predictors, a one-to-one mapping can be established between models with different error structures. To be more precise, consider a model where the outcome, $Y_t$, can be represented as a linear function 
of a scalar predictor,  $X_t$,  measured on day $t$ plus an error term with an AR(1) structure:

\begin{equation}
Y_t = \beta_0 + \beta_1 X_t + \eta_t,
\label{simpleAR}
\end{equation}

\noindent where $\eta_t = \phi \eta_{t-1} + \epsilon_t$ and $\epsilon_t$ is independent random error.  Then simple algebra establishes that 

 \begin{equation}
 Y_t = (1-\phi)\beta_0 + \beta_1 X_t + \phi Y_{t-1} - \phi \beta_1 X_{t-1} + \epsilon_t.
 \label{simpleARvialags}
 \end{equation}
rquation (\ref{simpleARvialags}) implies that the AR model (\ref{simpleAR}) is equivalent to a standard linear regression model that includes not only $X_t$, but also
 $X_{t-1}$ and $Y_{t-1}$ as predictors,  but with some constraints on the coefficients. One could of course simply fit $Y_t = \beta_0 + \beta_1 X_t + \beta_2 Y_{t-1} + \beta_3 X_{t-1} + \epsilon_t$ and this would result in a slightly more general model, with the AR(1) as a special case. This equivalence suggests that if our only interest were in predicting future values of $Y_t$ based on past history, then using the multiple linear model
 (\ref{simpleARvialags}) with the extra lagged predictors should give essentially the same prediction as model (\ref{simpleAR}), albeit with slightly more uncertainty because of not accounting for the constraint required to give the two models exact equivalence. The equivalence between formulations   (\ref{simpleAR}) and 
 (\ref{simpleARvialags}) is actually somewhat encouraging for contexts like our water modelling problem where we believe that lagged values of rain and climate variables are likely to be important.  The equivalence suggests that allowing for a flexible error structure can compensate, to some extent, for a misspecified mean model.  We will explore this in more depth later in the paper via simulations.  Further discussion can also be found at 
 \url{https://robjhyndman.com/hyndsight/arimax/}.

\subsection{Neural network modelling} 

The past several decades have seen an explosion of interest in the class of machine learning approaches known as neural networks.   Chapter 11 of  \cite{HTF} gives a very readable introduction that holds particular appeal for statisticians because it is grounded in familiar statistical modelling principles. It also shows how the approach relates to projection pursuit regression \citep{FriedmanStuetzle}.  Chapter 18 of \cite{efron2016computer} also provide
a clear overview that will appeal to statisticians.  
Drawing heavily on the material in both these sources,
 we start off in this section by briefly reviewing the basic concept of neural network modelling before then discussing the various extensions to accommodate data that are indexed in time. 
 
Neural network models are often referred
   to as {\it feedforward networks}, FFNN, because of the way that constructed functions of the inputs are passed through the hidden layer(s) to the output.  Another commonly used term is
   {\it multilayer perceptron}, MLP. The learning process of the FFNN consists of minimising the difference between the model output and the real data through the iterative adjustment of a set of internal weights. The updating process is performed via the {\it back-propogation} method which is well described by \cite{lecun2015deep} as well as by \cite{HTF} and  \cite{efron2016computer}.
   
\begin{center}
 \begin{figure}[!ht]
 \centerline{\includegraphics[ width =3.5in, height=2.7in]{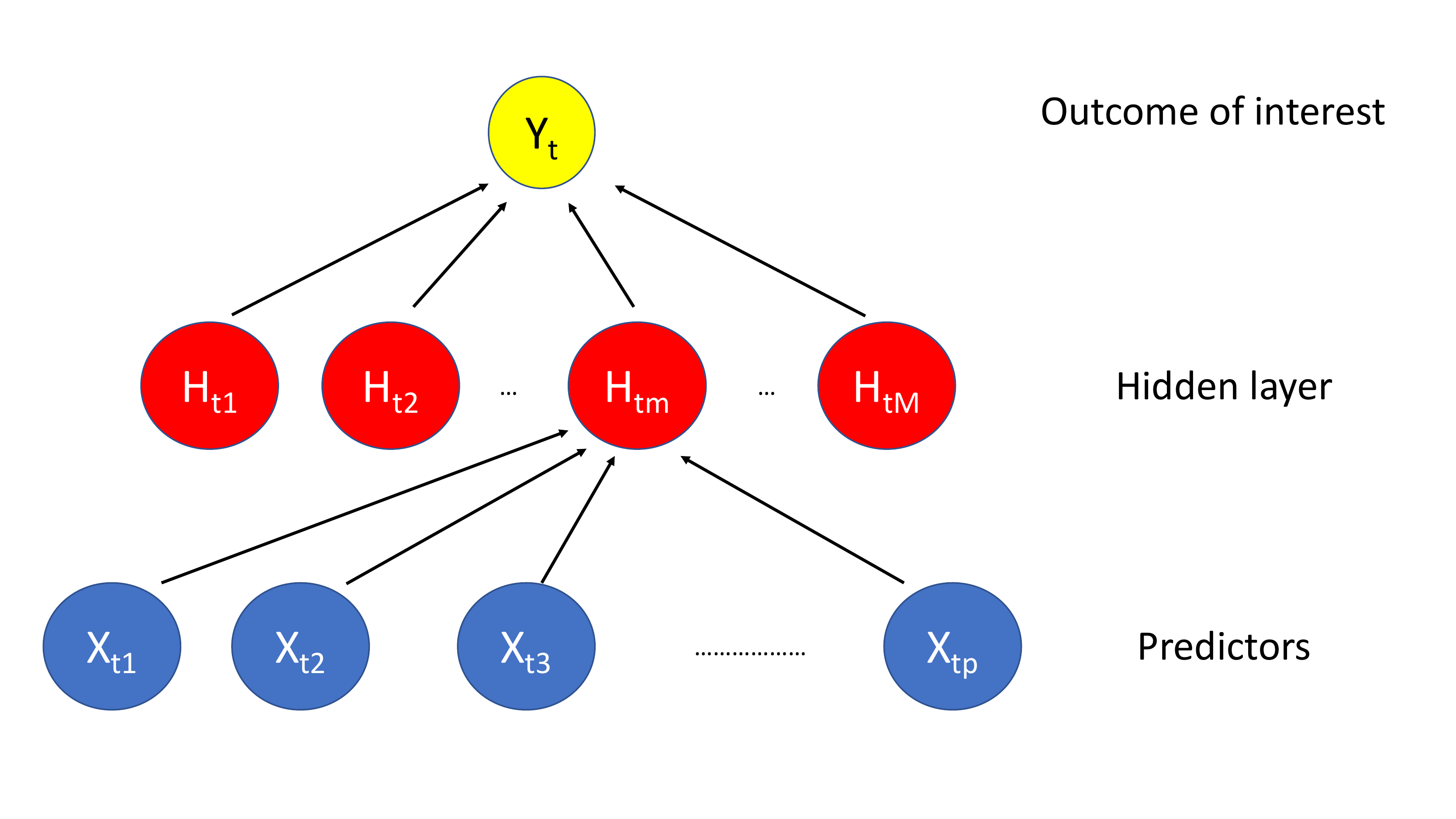}}
\caption{Neural network model (FFNN) with one hidden layer}
 \label{NNsinglelayer}
\end{figure}
\end{center}

   In general a neural network model works as follows. 
Suppose our goal is to predict a continuous outcome, $Y_t$, as a function of a set of predictors, $\mathbf{X}_t$ (for $t=1,...T$) with a neural network. To begin with, we will ignore  the time aspect and think of each $Y_t$ and $\mathbf{X}_t$ as an  independent pair. In contrast to a classical multiple linear regression model that might model the mean of $Y_t$ as a linear function of the elements of the predictor vector, $\mathbf{X}_t$, 
a neural network model creates a set of new variables, 
 $\mathbf{H}_t = \{H_{tm}\}_{m=1,...M}$, each of which is a potentially non-linear function of the elements of   
 $\mathbf{X}_t$.   In contrast to \cite{HTF} who use the notation $Z$ for these new variables, we use $H$ since this helps as a reminder that these 
 variables are part of 
a hidden layer and also because $H$ is commonly used notation in the machine learning literature. Figure \ref{NNsinglelayer} depicts a simple single layer neural network model.
The outcome of interest, $Y_t$,
 is then predicted as a function, $\sigma_y$, of
 a linear combination of the of $H_{tm}$ elements: 
 \begin{equation}
   Y_{t} = \sigma_y(\beta_{0} +  \bbeta_{1}^T \mathbf{H}_t), 
   \label{Ypred}
   \end{equation}
where $\sigma_y$ is referred to as an activation function (a nonlinear mapping function). In theory a wide range of different functions might be chosen for this output step, but in the case of a continuous outcome, a linear activation function (where no transform is applied) is 
 typically  used. 
 
 While it is common in the machine learning framework to think in terms of
finding good prediction models, using the prediction model in (\ref{Ypred}) with a linear activation function
can be
equivalently thought of in a more classical statistical framework as fitting 
a multiple linear regression model: 
 \begin{equation}
   Y_{t} = \beta_{0} +  \bbeta_{1}^T \mathbf{H}_t + \epsilon_t, 
   \label{Y}
   \end{equation}
where $\epsilon_t$ is a random error.  

The step involving the mapping from the $\mathbf{X}$-space to  the $\mathbf{H}$ space is critical since it is here that the modelling
 process potentially gains the flexibility to incorporate non-linearities and interactions among the elements of $\mathbf{X}$.  A typical model for the $m$th unit in the hidden layer is
 
 \begin{equation}
   H_{tm} = \sigma_h(\alpha_{0m} +  \balpha_{1m}^T \mathbf{X}_t), 
   \label{H}
   \end{equation}
   where $\sigma_h(\nu)$ is another activation function.  
   While a  wide range of functions could  be used here, 
   the sigmoid function
($\sigma_h(\nu) = 1/(1+e^{-\nu})$ ) or hyperbolic tangent (tanh) function 
($\sigma_h(\nu) = (e^{\nu}-e^{-\nu})/(e^{\nu}+e^{-\nu})$ )
 have traditionally been used for single-layer networks. Recently, however, a more popular choice has become the
  {\it rectified linear unit} or {\tt RELU} function which takes the form
  $\sigma_h(\nu) = \nu$ if $\nu>0$, 0 otherwise.  The {\tt RELU} function is now the most-used (and default) activation function for FFNNs as it has some computational advantages when used with multi-layer networks, which we will discuss later. 
  In our simulation and application sections, we will be fitting neural network models using the 
  R package {\tt keras} which offers these options and others.   
  
  As described by \cite{HTF}, we can think of the elements of $\mathbf{H}_t$ as a basis function expansion of the original inputs, $\mathbf{X}_t$ and therefore the neural network can be thought of as a standard linear model using these transformed variables.   Similar concepts have been discussed more recently by \cite{Cheng}.
 
To fit a neural network model, one must specify a 
   loss function, $R(\btheta)$, to measure how well the model predictions fit the actual data, where    $\btheta$ denotes the unknown $\bbeta$ and  $\balpha$ coefficients  from
   (\ref{Y}) and (\ref{H}). 
   For a continuous $Y$,  it is  common to use a loss function of mean squared error: 
   \[ R(\btheta) = \sum_{t=1}^T (Y_t-\hat{Y}_t)^2/T\]
   or 
   mean absolute error: 
   \[ R(\btheta) = \sum_{t=1}^T |Y_t-\hat{Y}_t|/T,\]
   where $Y_t$ is the observed value and $\hat{Y}_t$ is the
   predicted value at time $t$ based on inputs
    $\mathbf{X}_t$ and the unknown parameters $\btheta$.  (Note that 
   the machine learning literature often refers to the elements of $\btheta$  as weights rather than parameters.)   After specifying
   appropriate starting values for $\btheta$ (eg. by initialising weights close to zero, or setting them randomly), estimation proceeds through successive optimisation steps designed to
   reduce the loss function $R(\btheta)$.  As described in both  \cite{HTF} and \cite{efron2016computer}, gradient descent works particularly well in the
   context of neural networks and is more computationally manageable than approaches such as Newton-Raphson
   that are commonly seen in the statistical literature.  
   
Broadly speaking, these fitting steps are are not terribly different from traditional statistical model fitting.   In particular, if mean squared error is used for the loss function, the procedure
can be thought of as fitting a multiple linear regression predicting $Y$ as a function of the elements $\bH$, which can in turn be thought of as new predictors that have been engineered from the original input vector, $\bX$.  

 The term {\it deep learning} usually refers to extensions of the simple
  single-layer model in Figure \ref{NNsinglelayer} to include additional hidden layers, each of which is characterized as a potentially non-linear function of the units in the previous layer. While \cite{HTF} only discuss
  single-layer neural networks, \cite{efron2016computer} discuss the more
  general multi-layered case.  

A serious challenge with neural network models is that they tend to be highly over-parameterized, 
especially when there are multiple layers with many units per hidden layer and many potential predictors.  
For example, a 3-layer neural network with $M_1, M_2$ and $M_3$ hidden units each will have 
$M_1(p+1) + M_2(M_1+1) + M_3(M_2+1) + M_3+1$ parameters, 
where $p$ refers to the dimension of the 
predictor vector $\mathbf{X}$.  Thus, careful strategies are needed to avoid over-fitting.  

One approach is to use {\it regularization} or {\it penalization} of the $\balpha$s and $\bbeta$s with either $L_1$ or $L_2$ norms, just as we might do when applying a technique such as Lasso or ridge regression in classic regression
settings where the numbers of predictors are very large relative to the sample size \citep{tibshirani}.  
 In the machine learning world, the term {\it weight decay} refers to use of the $L_2$ norm to regularise  the parameter estimates \citep{chollet2018deep}. While a penalisation approach is appealing  because of its strong grounding in traditional principles of
statistical inference, the approach can be computationally demanding as a result of the more complicated objective function. In practice, it is common to instead use a 
so-called {\it early stopping rule} approach or {\it dropout}, which works as follows. 

A neural network is trained in a series of {\it epochs}, or iterations, during which the data are shown to the model in batches of pre-specified size with an updating step occurring with each batch. At the end of each epoch (after the entire data set has been seen by the model), the loss function $R(\btheta)$ is calculated using both the training data set and the validation data, based on the current estimated  values of $\btheta$. This iterative process is repeated until there is evidence that the algorithm is starting to overfit the training data and losing the ability to reliably predict the validation data. This generally corresponds to the point where the loss function evaluated on the validation data set starts to increase, whereas the loss function evaluated on the training data set continues to decrease. A plot of the loss functions for the training and validation data is useful for finding this point, or an automated early stopping method can be incorporated into the training procedure.

Early stopping works by attempting to halt the gradient descent algorithm before convergence of the model, at a point that represents the best generalisation potential. The loss calculations that take place after each epoch are used for this. The loss function on the validation data is monitored until it fails to decrease for a specified number of epochs, with this number known as the {\it patience}, indicating a point where improvements of the model based on the training data are no longer improving its representation of the validation data set, and therefore the ability of the model for generalisation is decreasing. The early stopping technique is extensively explained by \cite{Goodfellow-et-al-2016}. There can be something of an art in deciding when to stop training a model, and it is generally a good idea to try a few different strategies and compare the resulting model fits.  We will see this quite clearly with the simulations described in the next section.

Dropout is another technique used to reduce overfitting, popular due to its simplicity of implementation. It works by reducing the dependence of the model on any single node, by essentially creating an ensemble of models with fewer nodes and varied model architectures. Nodes are randomly dropped out of the model for each training run to create these subnetworks by means of randomly deactivating a proportion of the nodes (typically 10-50\%) through setting their activation to zero. All nodes are reactivated for the prediction phase.

 \cite{Goodfellow-et-al-2016} provide a chapter explaining these and other regularisation techniques for deep learning in great detail. 
 
 In addition to trying a range of strategies to decide when the training of a model has reached a good stopping point, there are other options that an analyst can vary, in something of a sensitivity analysis, to improve the fit of neural networks. These steps, collectively, are known as tuning the hyperparameters of the model.  For example, it will make sense to explore whether the model fit can be improved by adding extra layers, adding more units per layer or changing activation functions. We will have more discussion and illustrations of how these optimization and tuning steps work presently. 
 
 Using the same validation data to decide on the best prediction runs the risk of overfitting again.  So in practice, it is common to divide the data set into three portions:  training data used to run the optimization algorithm, validation data to decide on the appropriate model specifications (stopping time, number of layers, number of units per layer, etc) and finally, the test data to measure the overall prediction quality for the final tuned model.

\subsection{Neural networks for the time series setting}

In this section, we consider some of the strategies that have been proposed
in the machine learning literature 
over the past several decades to generalize  neural network methodologies to accommodate data that are indexed by time. The simplest approach builds on the duality that we touched on briefly in subsection \ref{classical}, namely that classical time series regression models can generally be re-expressed 
as standard linear regression models that build a richer predictor space by including lagged values of
predictors and outcomes.  

To make our discussion clearer, we link  back to a simplified version of
our motivating application where
the purpose is to build a model that predicts bore water levels as a function of rainfall.  Let  $Y_t$ represent the
water level measured in the bore at time $t$ and let $rain_t$ be the rainfall measured near the site on day $t$.
Now consider the following predictor vector: 
\[ \bX^T = \left(rain_t, rain_{t-1}, rain_{t-2}, \dots, rain_{t-k}, Y_{t-1},Y_{t-2}, \dots, Y_{t-l}\right). \]
\cite{MakridakisEtal} considered a similar model, though only including the lagged $Y$ variables, in a multilayer perceptron. They used cross-validation to decide on the optimal number of lagged $Y$ values to include.  While this approach has some appeal due to its simplicity, there are some limitations in practice.  First of all, while it is straightforward to use for 1-step ahead forecasting, extensions beyond this are more challenging since the required
predictors will not be available.  Another problem is that models requiring many lags can become unstable, especially in the presence of high autocorrelation in the predictor space.  In our motivating application, it is common to find that water levels can be predicted by rainfall levels over the past several months, and therefore the model would require over 60 embedded lags, possibly leading to instability.   

{\it Recurrent neural networks (RNNs)} were proposed back in the 1990s specifically for temporal data, as a means of addressing such problems. With RNNs, data is fed through sequentially and the hidden layer weights are shared across all the timesteps. Each output becomes a function of the predictor values at previous times in the timeseries. 

\cite{Goodfellow-et-al-2016} describe
how the back-propogation  algorithm used for fitting feedforward networks is extended to the recurrent network context, via {\it back-propogation through time} where the principle is similar to that used in standard neural network models. They explain that RNNs can be thought of as feedforward networks that have been extended to include feedback connections. Much of the early developments in RNNs were motivated by the challenge of developing
automated ways for computers to read and interpret text data.  The idea was that good algorithms to interpret the meaning of a sentence 
should not just base the analysis on looking word by word, but rather consider
each word in the context of the sequence in which it appears. 

\begin{figure}[H]
    \centering
    \includegraphics[ width =3.1in, height=2.2in]{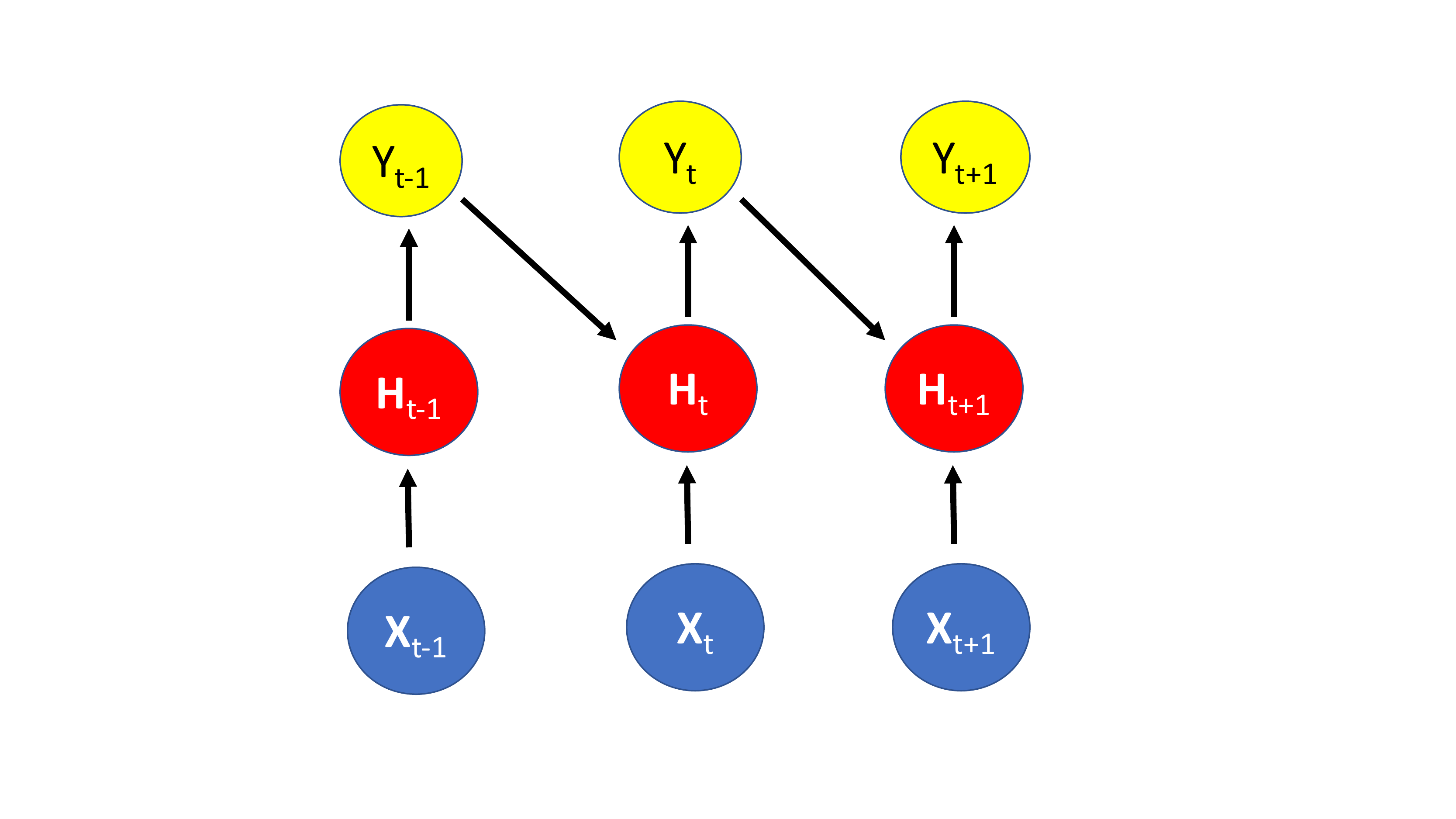}
    \includegraphics[ width =3.1in, height=2.2in]{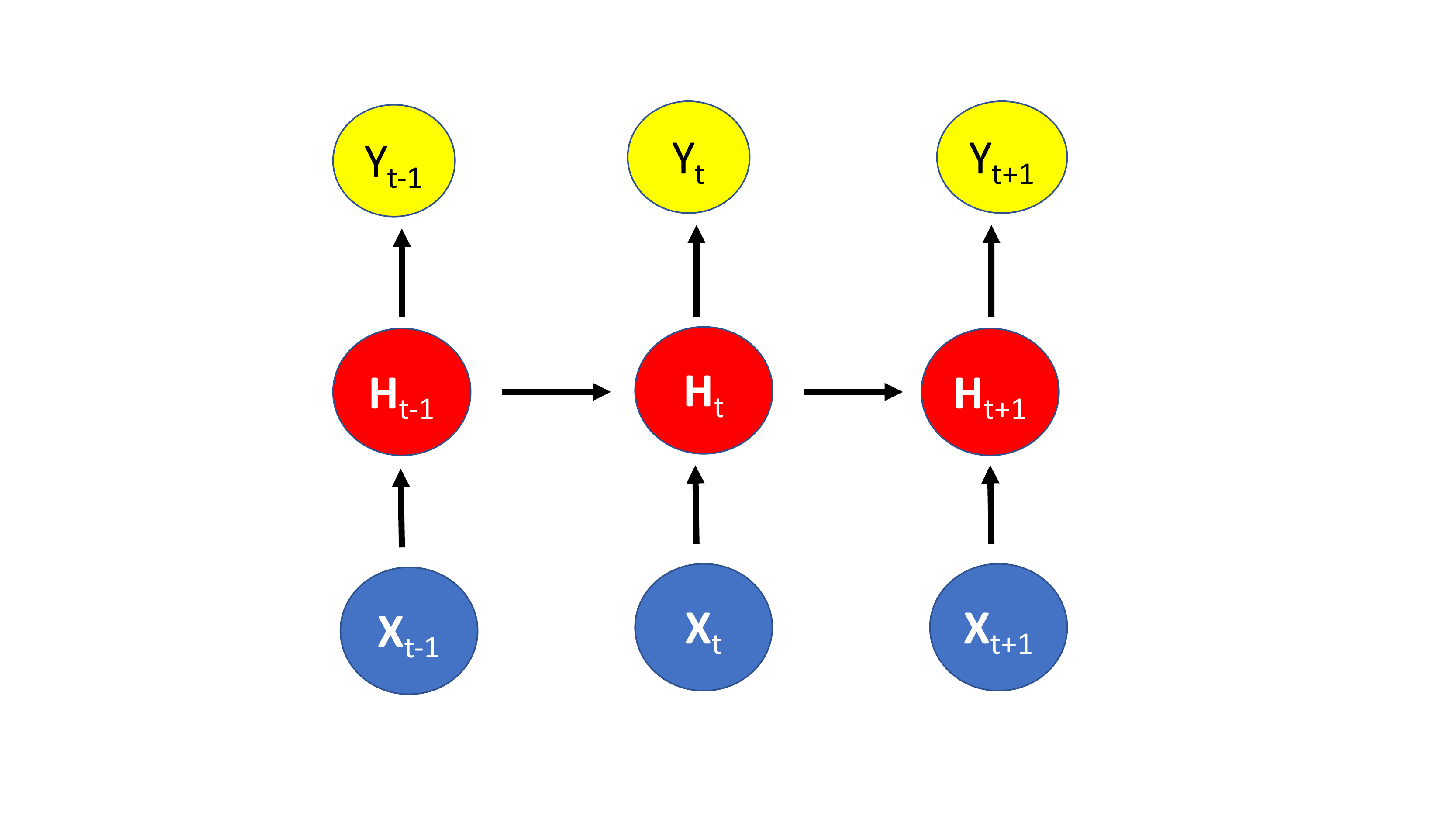}
    \caption{Simple Jordan (left) and Elman (right) recurrent neural networks}
    \label{fig:RNNs}
\end{figure}

Indeed, two of the  earliest
neural networks for sequence data were developed in the context of research into cognition \citep{elman} and speech  \citep{Jordan}.  The basic idea is that instead of explicitly
including lagged values of $Y$ as part of the predictor space, one instead includes lagged values of either the predicted $Y$s or the estimated hidden layers. These models are depicted in Figure \ref{fig:RNNs}.  

In a simple Jordan recurrent neural network (left panel of Figure \ref{fig:RNNs}),
the outcome $Y$ is predicted just as in (\ref{Ypred})
as a function of the elements of a hidden layer, $\bH_t$, 

\begin{equation}
    Y_t = \sigma_y\left(\beta_0 + \bbeta_1 \bH_t\right),
\label{Jordan_Y}
\end{equation}
but the expression for the hidden layer neuron $m$ at time $t$, corresponding to Equation (\ref{H}),  now includes a lagged term: 
\begin{equation}
   H_{tm} = \sigma_h\left(\alpha_{0m} +  \balpha_{1m}^T \mathbf{X}_t + \alpha_{2m} Y_{t-1}\right). 
 \label{jordan_H}
 \end{equation}
 
It is important to note that in this last expression for $H_{tm}$, it is not the observed value of $Y_t$ that is used, but rather, the prediction from (\ref{Jordan_Y}). When $\sigma_y$ and $\sigma_h$ are both  linear activation functions, then the prediction for the $m$th element of the hidden layer
can be expressed as
\begin{equation}
 H_{tm} = \alpha_{0m}^* +  \balpha_{1m}^T \mathbf{X}_t +   \balpha_{2m}^*\bH_{t-1}. 
\end{equation}
where $\alpha_{0m}^* =\alpha_{0m} +\alpha_{2m}\beta_0 $ and
$\balpha_{2m}^* = \alpha_{2m}\bbeta_1$. 
This expression has an appealing link to the concepts of autoregressive models discussed in the previous section.  A simple Elman layer \citep{elman}, depicted in the right hand panel of Figure \ref{fig:RNNs},  replaces (\ref{jordan_H}) with

\begin{equation}
   H_{tm} = \sigma_h\left(\alpha_{0m} +  \balpha_{1m}^T \mathbf{X}_t + \balpha_{2m} \bH_{t-1}\right). 
 \label{elman_H}
 \end{equation}
so that the individual nodes' values from the hidden layer at the previous timepoint enter as predictors of the node values at the current timepoint.

Chapter 10 of \cite{Goodfellow-et-al-2016} provides a good discussion on these two modelling strategies,
arguing that networks that allow hidden-layer-to-hidden-layer connections as in (\ref{elman_H}) give more flexibility, but at the cost of significantly more computational effort.   

Since the Jordan and Elman recurrent neural networks were proposed in the early 1990s, there have been many extensions to RNNs designed to capture more real-world complexities.  These include the addition of multiple, as opposed to a single, hidden layers as well as the incorporation of multiple time steps 

A more general RNN structure looks the same as the Elman network in Figure \ref{fig:RNNs} but with as many time steps as desired added to the lefthand edge. 
In the RNN, the layer weights remain the same as the time series cycles through, and each output element depends on previous values of the predictor time series.

As people started exploring the use of recurrent neural networks with multiple time steps they found that, just as described above for the multilayer perceptron, the fitting could easily become unstable. In particular, the method does not tend to work well in settings where it might be important to incorporate longer-term memory.   The term {\it vanishing gradient problem} was created to described situations where instabilities arise when the algorithm needs to accumulate a large number of very small derivatives in order to take an optimisation step. The training signal passing through the network becomes exponentially small due to repeated multiplication. The weights on the long-term interactions then end up much smaller than the weights on the short-term interactions, making it more difficult to identify correlations between events that are more separated in time.  

The {\it Long Short-Term Memory} or LSTM model was proposed in the late 1990s as a solution to the vanishing gradient problem  \citep{hochreiter}.  The LSTM is an updated RNN  involving the inclusion of a series of 'gates' that regulate the transfer of information through time, and memory cells that store information from the past. The gates are termed the {\it input} gate, the {\it forget} gate and the {\it output} gate. These determine when to carry information forward through time and when to forget it, allowing important information from the long past to be retained in the memory cells whilst non-essential information is discarded. The input, forget and output gates, for the $m$th memory cell are modeled respectively as:
% \begin{equation}
\begin{align}
   i_{tm}& = \sigma\left(\alpha_{0mi} +  \balpha_{1mi}^T \mathbf{X}_t + \balpha_{2mi} \bH_{t-1}\right)
    \label{LSTM_i}\\
     f_{tm}& = \sigma\left(\alpha_{0mf} +  \balpha_{1mf}^T \mathbf{X}_t + \balpha_{2mf} \bH_{t-1}\right) \\
   o_{tm}& = \sigma\left(\alpha_{0mo} +  \balpha_{1mo}^T \mathbf{X}_t + \balpha_{2mo} \bH_{t-1}\right). 
 \label{LSTM}
 \end{align}
%  \end{equation}
  where $\sigma(\nu)$ is the sigmoid function, providing a gating value of between 0 and 1 that regulates the information flow through each gate, and the $\alpha$'s are a separate set of learnable parameters for each gate. 
  
  The state of the memory cell, $s$, is updated based on the previous system state and the current input. A potential cell state update, $\Tilde{s}_{tm}$, is first proposed based on the current input:
    \begin{equation}
  \Tilde{s}_{tm} = \tanh\left(\alpha_{0m} +  \balpha_{1m}^T \mathbf{X}_t + \balpha_{2m} \bH_{t-1}\right)
   \label{LSTM_stilde}
 \end{equation}
  and with regulation by the forget and input gates, the memory cell state is updated as:
   \begin{equation}
   s_{tm} = f_{tm}s_{(t-1)m}+i_{tm}\Tilde{s}_{tm}.
 \label{LSTM_s}
 \end{equation}
 The elements of the forget and input gates determine which components of the past system state and current input will be used to update the current system state (those with $f_{tm}$ and $i_{tm}$ closer to 1) and which will be discarded (those with $f_{tm}$ and $i_{tm}$ closer to 0).
 
 The memory cell output is calculated using the updated system state with tanh activation, regulated by the output gate:
   \begin{equation}
   h_{tm} = \tanh(s_{tm})o_{tm}.
 \label{LSTM_o}
 \end{equation}
 The output gate controls the flow of information from the cell state to the hidden layer. Use of the tanh activation provides output from the memory cell in the range [-1,1]. 
 
 Corresponding to \ref{Ypred}, the predicted model output takes the form:
    \begin{equation}
   Y_{t} = \sigma_y\left(\beta_{0} +  \bbeta_{1}^T \mathbf{H}_t\right),
 \label{LSTM_y}
 \end{equation}
 where again, $\sigma_y$ is usually the linear activation function in regression problems.

The LSTM overcomes the vanishing gradient problem encountered with traditional RNNs and allows for much longer time series to be analysed. Multi-scale time dependencies can be found with the LSTM whilst incorporating forcing data. For a detailed description of the LSTM model structure, see \cite{vlachas2018data}, \cite{greff2016lstm} or \cite{kratzert2018rainfall}.

In the next section, we will use simulations to evaluate the performance of the various methods that have been described in this section.  All analyses were conducted in R version 3.6.3 \citep{R}, with packages {\tt keras} \citep{Keraspackage} for neural network modelling  and {\tt fable} \citep{fable} for ARIMA modelling. \cite{chollet2018deep} and \cite{ghatak2019deep} are good resources for guidance on fitting neural network models with R. While all models could be run on standard laptops, some of the neural network models could take several hours to run.  Use of a high performance cluster allowed us to efficiently conduct simulations and also to tune the models, which required running them over a grid of modelling options.  This will be discussed in depth in the next section.

\section{Simulations}
\label{simulations}

In this section, we construct simulation studies designed to approximately mimic the behavior of groundwater recharge as it responds to a series of rainfall events. We  then investigate various approaches to analyzing the resulting datasets using classical models as well as various types of neural networks.

The hydrogeological literature is replete with
articles discussing the complex relationships that exist within groundwater systems. Climatic conditions (such as rainfall, temperature, evaporation and humidity), and physical surface and subsurface attributes (such as vegetation cover, soil characteristics, and subsurface geological structures) are among the many factors  associated with local aquifer water levels.  
Developing a simulation that realistically captures the complexity of these relationships is beyond the scope of this paper and is not our goal. Many physically-based models already exist that attempt to represent the complex interactions of groundwater systems by means of comprehensive sets of differential equations that characterise the interactions between each component of the system. Our aim, rather, is to create some simulations  that are  effective in capturing enough of this complexity to form the basis for a helpful comparison between the various methods that were presented in the previous section. 

We begin by focusing on predicting aquifer levels as a function solely of rainfall, while allowing for the possibility of some additional long-term trends which may exist, for example, due to a gradual reduction in water levels due to extractions or drought. Following this simple simulation, we perform a more complex simulation adding evapotranspiration as an additional time-indexed predictor and increasing the non-linearities in the intervariable relationships.    We do not simulate rainfall
data directly, but rather use a block bootstrap to generate multiple sequences from
an actual multi-year record of rainfall. 

\subsection{Climate data}

\begin{figure}[H]
\centering
\includegraphics[width=.8\columnwidth, height=3in]{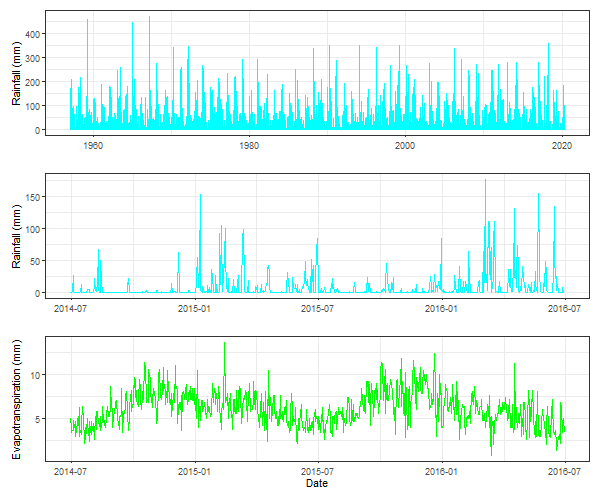}
\caption[short title]{Observed daily rainfall and evaporation at station 32042 (Tully, NSW, 1957-2020). All available rainfall data (63 years) are shown on the top plot, and two years of seasonal patterns in rainfall and evaporation are evident on the lower plots.}
\label{fig:obsTully}
\end{figure}

The rainfall timeseries in our simulations  are generated from actual rainfall measurements at Tully, a town in the north eastern part of the Australian state of Queensland. Sixty three years of recorded data from this station (1957-2020) were obtained from the Queensland Department of Environment and Sciences  SILO database for meteorological variables (ref: https://www. longpaddock.qld.gov.au/silo/, station 32042). Tully is informally recognized as being one of the wettest places in Australia and its location in the tropical far northeast of Australia gives it a strong wet/dry seasonal pattern. 
Consequently we felt that it would be a good candidate for generating a strong
pattern with bore water levels in our simulation
models.  The Tully data (rainfall as well as evapotranspiration) are shown in Figure \ref{fig:obsTully}.

To produce multiple 10-year long timeseries for use in  simulation modelling, we repeatedly draw bootstrap samples from the measured Tully data using the
double seasonal block bootstrap method, described \cite{HyndmanFan}.
This method builds datasets from blocks of variable-length, consecutively measured observations extracted from the original data, which are placed in the new datasets within a specified number of days from the date of the year on which they were observed. Seasonality in the data is preserved without requiring blocks to be an entire year in length, as the observations are moved randomly between years but stay roughly at the same time of year as they were measured. We use blocks of an average 60 days in length, with a maximum $+/-$ 14 days variation in the start date. An autocorrelation plot (not shown) indicates that blocks of 60 days will capture most of the autocorrelation present in the Tully data.

\subsection{Simulating groundwater levels - simple model} 

\begin{figure}[H]
\centering
\includegraphics[width=.5\columnwidth]{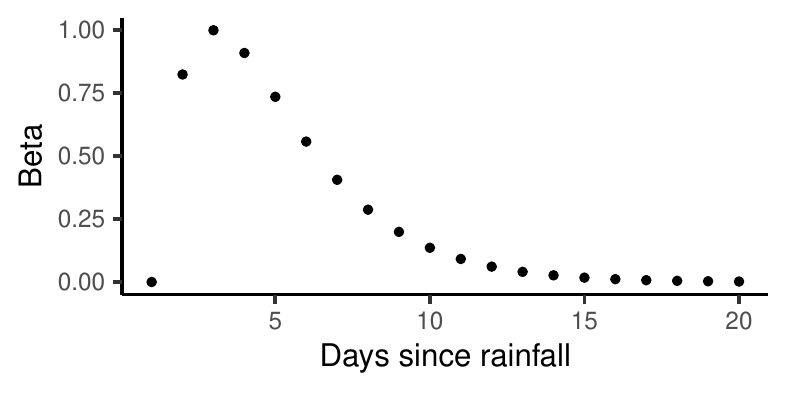}
\caption[short title]{Influence of rainfall on aquifer level in simulation data, by days since rainfall occurred.}
\label{fig:influencerain}
\end{figure}

For each bootstrapped rainfall data set, we generate groundwater levels  following the logic of  \cite{BakkerSchaars}. They  assumed
 that rain falling on a particular day $t$ will continue to impact the water level in the aquifer  for an additional number of days, which we define as $t_{mem}$, indicating that this is how many days of rainfall {\it memory} will be needed for prediction.   As suggested by \cite{BakkerSchaars}, we assume that the coefficient relating rainfall on day $t-s$ to the predicted aquifer level on day $t$ follows the shape of a scaled gamma function. 
 More precisely, we assume that $Y_t$, the level of water in the aquifer on day $t$, can be expressed as

\begin{equation} 
Y_t = \beta_0 +     \sum\limits_{s=1}^{t_{mem}} rain_{t-s} \beta_s + \epsilon_t, 
\label{BSModel}
\end{equation}
where $\epsilon_t$ represents an error term to be discussed presently,   $rain_{t-s}$ corresponds to the amount of rain on day $t-s$ and
\[\beta_s \propto  (s-t_{mem})^{a-1} \exp\left[-(s-t_{mem})/b\right]. \] 
   The weighting function for the influence of the previous days' rainfall on groundwater levels is shown in Figure \ref{fig:influencerain}. For our simulation, we have chosen the values of $a$ and $b$ to ensure that the effect of rainfall becomes negligible after 20 days (ie. $t_{mem}=20$).

To generate some   non-linearity in  the groundwater response, we incorporate some simple modifications  into equation (\ref{BSModel}),  motivated by concepts of evapotranspiration and soil moisture in the rainfall/infiltration physical system. Rainfall amounts in the lower 40th percentile of all non-zero events (approximately those less than 6mm/day) are set to zero, to represent the interception of small amounts of rainfall by vegetation and the initially dry conditions of the surface soil layer. Large rainfall events, greater than the 95th percentile (approximately those from 101-450mm/day), are set to reach only a quarter of their original magnitude above 100mm, to characterise surface runoff that occurs when the soil is fully saturated. 

The error terms in (\ref{BSModel}) are generated from an autoregressive fractionally integrated moving average (ARFIMA) model using the {\tt arfima} package, which includes a particularly simple and convenient simulation function. 
The ARFIMA model is similar to the ARIMA model (\ref{ARIMA}) discussed earlier, except that $d$ can take values other than 0, 1 or 2. In our simulation, we use $p=2$ with $\phi = (0.4, 0.2)$ (AR parameters), $d = 0.4$, $q=1$ with $\theta = 0.5$ (MA parameters).

\begin{figure}[H]
\centering
\includegraphics[width=.8\columnwidth]{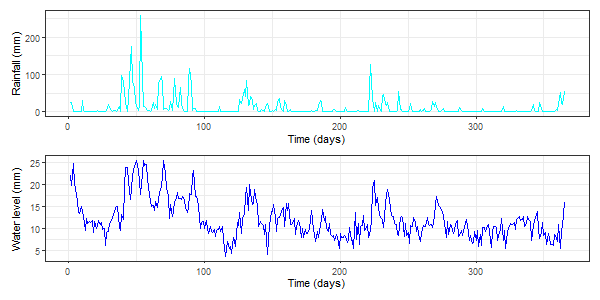}
\caption[short title]{A sample year of generated rainfall and water level timeseries.}
\label{fig:generateddataset}
\end{figure}

A sample dataset in which the response of groundwater level is a function of the weighted, lagged, cropped rainfall with ARFIMA error structure is shown in Figure \ref{fig:generateddataset}, with the top panel showing a bootstrapped set of rainfall data and the lower panel the generated bore level measurements.   

We fit a series of models, ranging in complexity, on each of the generated data sets. The models include a classical timeseries model (ARIMA), 
three classic neural networks (a linear FFNN, a one-layer FFNN, a two-layer FFNN), and finally a one layer and a two-layer recurrent neural network (LSTM). 

For training the neural networks, the data are split into training, validation and testing sets comprising 60\%, 20\% and 20\% of the sequential data respectively. The models are created with the training set, and validated during the training process with the validation set. Reported errors are calculated by comparison of predictions using the testing set predictors to the true testing set response values. The ARIMA model does not expect a validation set, and so the split for ARIMA is 80\% training data and 20\% testing data.

The ARIMA model requires explicit incorporation of
lagged values of predictor variables from previous timesteps. This is easily accomplished using the  {\tt embed} function in {\tt R} and once done, running the ARIMA function from the {\tt fable} package is straightforward. 

Lagged predictor variables are also incorporated explicitly into the three FFNN models. For now, we consider only models
incorporating the correct number of lags which, based on our simulations, we know to be 20.  It is not necessary to explicitly incorporate
lagged predictors into the LSTM models since this is handled as part of the modelling
process.  

In  contrast to ARIMA modelling, running the neural network models involves quite a lot of 
setup in terms of standardizing the input data, 
determining an appropriate loss function for use in network training, and specifying the choice of activation functions used in the transfer of data between layers of the network.  

Standardising each variable into a range between 0 and 1 is customary to ensure each variable receives a similar representation by the neural network. This is especially important when variables are measured in different units with differing measurement scales. The scaling of the entire data set is based on the data in the training set, with the same scaling applied to the validation and testing sets. This means that the training data will always be in the range [0,1] but the validation and testing data will not necessarily reach either limit, or may in fact be outside the range, depending on the values of observations in the validation and testing sets compared with the observations within the training set.

 We chose mean square error for the loss function, though we also monitor mean absolute error for comparison.  

Our simulations use the {\tt RELU} activation function in the hidden layers of the 1 and 2-layer FFNNs, but a linear activation function for the 'linear FFNN' model.  The latter means that no transformation occurs and this model effectively reduces to ordinary linear The LSTM traditionally incorporates inbuilt sigmoid and a hyperbolic tangent activation functions within the memory cell, and we have not altered these defaults. Due to the data passing through these nonlinear transformations within the memory cell, it is common practise to not include any further activation at the nodes of the hidden layers. There is, however, discussion in the literature about choosing different activation functions for RNN's, for example \cite{farzad2019comparative} investigate alternatives to the sigmoid activations at the LSTM input, forget and output gates. We add an activation at the final output layer only, and as our task is regression we use a linear activation here as we do in the FFNN.

\begin{figure}[H]
\centering
\includegraphics[width=.6\columnwidth]{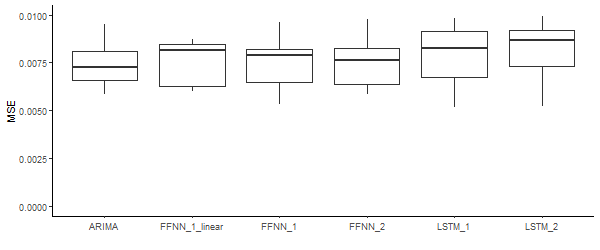}
\caption[short title]{Mean squared error on testing data sets for a variety of model types. Data include 20 lagged predictor values, and the neural networks have 32 nodes per layer.}
\label{fig:errors_simB}
\end{figure}

Figure  \ref{fig:errors_simB} shows boxplots of mean squared errors evaluated on the test data sets for ten different simulations, as described above.   The neural networks all include 32 nodes per layer, and are run for up to 1000 epochs, or until there is no improvement in MSE over 10 training epochs (patience=10), with a batchsize of 32. The neural networks are not optimised at this point, and the number of nodes has been chosen arbitrarily. 

We can see that on this somewhat simple data set all models do reasonably well due to the relatively linear nature of the  inter-variable relationships. The ARIMA model is displaying an advantage over the other models on this data, presumably because it is the most parsimonious model and is not as affected by overfitting as the various neural network models since the
latter tend to contain many more parameters to be estimated.

\subsection{A more complex simulation} 

We now increase the non-linearity within the time series generation process to determine if a larger spread in model outcomes would occur with more complex data.

\begin{figure}[!ht]
\centering
\includegraphics[width=.8\columnwidth]{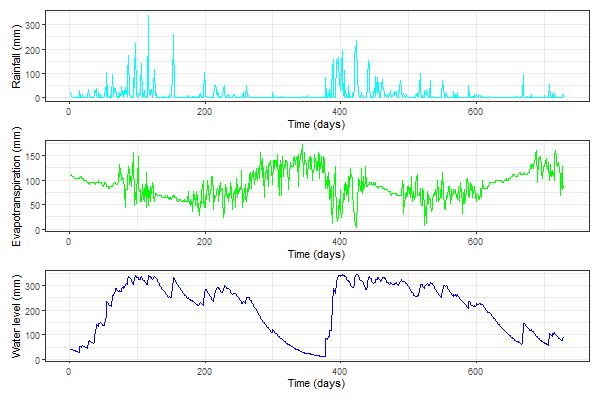}
\caption[short title]{Process-based simulation data - two years of generated rainfall, evaporation and groundwater level timeseries.}
\label{fig:gendata_process}
\end{figure}

A more physically-representative response time series is created with the GR4J hydrological model  \citep{perrin2003improvement}, an empirical model that routes inputs of rainfall and evapotranspiration through interception, infiltration, soil moisture storage, and streamflow components. Default parameters for the GR4J model are used (as described in https://wiki.ewater.org.au/display/ SD41/GR4J+-+SRG) with the bootstrapped climate data from Tully station as the forcing time series. On each day of the model simulation, a discrete model state is found after integration of the equations governing exchange between the components. We take the model state parameter representing soil moisture storage to symbolise our groundwater level response, as soil storage is closely related to groundwater head in a simple unconfined aquifer system. This process-based simulation dataset, shown in Figure \ref{fig:gendata_process}, exhibits relatively realistic non-linear dynamics for a simple unconfined aquifer responding to rainfall events and evaporation.

\begin{figure}[H]
\centering
\includegraphics[width=.6\columnwidth]{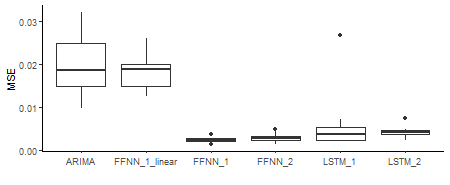}
\caption[short title]{Mean squared error on testing data sets for a variety of model types produced with 10 sets of generated data. Data include 50 lagged predictor values, batchsize = 32, patience = 10, and the neural networks have 32 nodes per layer.}
\label{fig:errors_simGR4J}
\end{figure}

As in Figure \ref{fig:errors_simB}, we check the MSE's for each model type when using the more complex simulated data, with results shown in Figure \ref{fig:errors_simGR4J}. The same parameters are used as for producing Figure \ref{fig:errors_simB} with the exception that we now include 50 lagged days of rainfall at each timepoint rather than 20, as this data was generated with a more complicated rainfall/water level relationship. We can see that with the more complex data set, the ARIMA and linear FFNN do similarly well, but the one- and two-layer FFNNs and LSTMs have consistently lower prediction errors.  The likely explanation 
is that the GR4J simulation has introduced significant non-linearities which the ARIMA and linear FFNN cannot capture. 

\subsection{Further tuning the LSTM}

The LSTM models can be further tuned by specifying appropriate choices of hyperparameters to allow the models to best represent the current data set. A number of hyperparemeters require specification while setting up the model, such as the width and depth of the network (number of nodes per layer, and number of layers), the number of embedded lagged inputs that will be included with each input observation, and the degree of regularization. There is currently no unique method for determining LSTM hyperparameters, as each choice produces a different modelling outcome and the influence of hyperparameters (and interactions between various hyperparameters) is difficult to quantify. Practices ranging from the use of values from previous projects, arbitrarily chosen values, or a grid search of the various possible combinations are all used in the literature for hyperparameter specification. 

 Some advice is given in the literature relating to these choices. \cite{Goodfellow-et-al-2016} suggest 
 that models with more layers tend to be more efficient and have greater potential for generalisation, but deeper models can tend to overfit if there are limited training samples.  \cite{greff2016lstm} found larger networks performed better, but with diminishing returns and increased computational costs. The number of epochs (iterations) of training data that the model sees during the training process is another important value to specify. This can be altered through the concept of 'patience' during the training procedure. When a patience value is set, training will continue as long as the MSE of the prediction on the validation data set continues to improve with each epoch. When this MSE fails to improve after the specified number of epochs, the model quits training.

The number of lagged inputs is a critical hyperparameter that needs to be considered when
using RNNs, as well as when running ARIMA models and the simpler FFNNs.  It seems that embedded lags are important in the LSTM when the \textit{sequence} of inputs is significant for learning long-term dependencies, in addition to when recent conditions impact the current situation. The true autocorrelation in the data can be of unknown length, as the LSTM incorporates its long-term memory of past system states to learn autocorrelations.

We have opted for the grid search method to evaluate our hyperparameter options, in which we have specified a discrete list of possible values for each hyperparameter, and run the models with each combination. This has led to 2870 combinations, which were each run with 10 simulated data sets on a high performance computing cluster. 

\begin{figure}[!ht]
    \centering
\subfloat[]{\includegraphics[width = 3.5in]{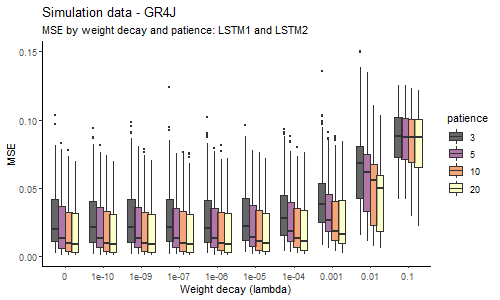}}
\subfloat[]{\includegraphics[width =3.5in]{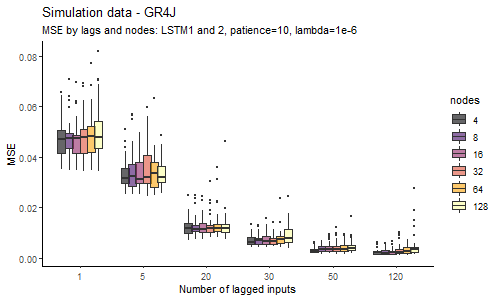}} \\ 
\subfloat[]{\includegraphics[width = 3.5in]{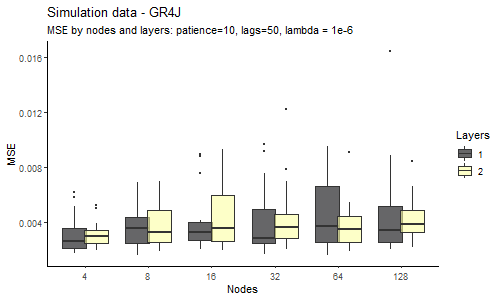}}
\subfloat[]{\includegraphics[width = 3.5in]{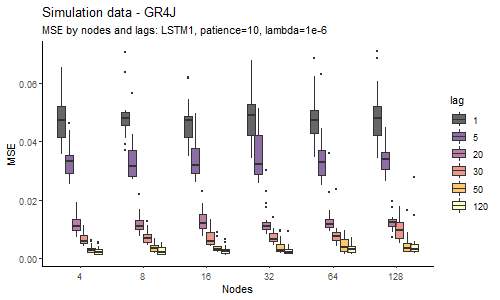}}
    \caption{The effect of hyperparameter choices on mean squared errors for LSTMs. In a) weight decay and patience parameters are compared; b) displays the relationship between lagged inputs and number of nodes for patience =10 and lambda = 1e-6; c) compares results with 1- and 2-layer LSTMs for the subset further refined to include only 50 lagged values; and d) allows the choice of nodes after the other parameters have been set. }
    \label{fig:LSTM1_hyp}
\end{figure}

Figure \ref{fig:LSTM1_hyp}  demonstrates the sensitivity of the LSTM to variations in the hyperparameters specified in the model setup. Boxplots of mean squared errors resulting from the 10 simulations are shown. We can see from subplot (a) that a patience value of 10 and a weight decay of $10^{-6}$ will be decent choices; from (b) we determine that including 50 lagged predictor values is sufficient; from (c) we see that a single layer generally produces the best results on this data; and from (d) we see that for the combination of parameters already chosen, 32 nodes will likely produce good results. These hyperparameters will be used in the model for prediction on the testing data set in the next section.

\subsection{Predictions with ARIMA and LSTM}

\begin{figure}[!ht]
\centering
\includegraphics[width=0.8\columnwidth]{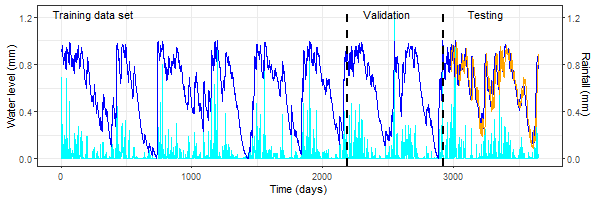}
\caption[short title]{Training, validation and testing data sets with LSTM predicted output.}
\label{fig:trainvaltest}
\end{figure}

We will now use the LSTM and ARIMA models that we have developed to predict the aquifer level based on observed rainfall and evapotranspiration. The predictions will be compared with the observed water levels in the testing portion of the dataset. Figure \ref{fig:trainvaltest} shows a generated data set split into the training, validation and testing sets. The prediction by LSTM is shown on the testing portion of the data, over the observed water levels. For the ARIMA model, both the training and validation sets shown here are considered the 'training' set.

\begin{figure}[!h]
\centering
\subfloat[]{\includegraphics[width=.8\columnwidth]{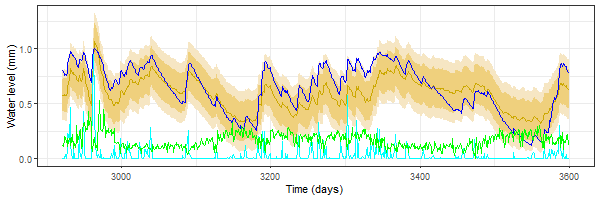}}\\
\subfloat[]{\includegraphics[width=.8\columnwidth]{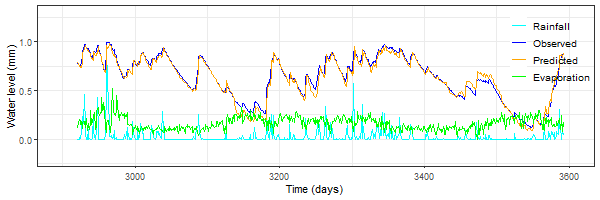}}
\caption[short title]{Predicted groundwater level time series, with a) ARIMA(1,0,1) prediction, with 50 lagged predictor values and d set to 0, showing 80 and 95\% prediction intervals (MSE=0.019); and b) a 1-layer, 32 node LSTM with 50 lags, 10 epochs of patience and weight decay parameter 1e-6 (MSE=0.002).}
\label{fig:modeloutput_GR4J}
\end{figure}

The predictions produced by the ARIMA and LSTM models are shown in Figure \ref{fig:modeloutput_GR4J}, which displays only the testing portion of the data set from Figure \ref{fig:trainvaltest}. The best-fitting ARIMA model was found to be ARIMA(1,0,1), when $d$ was set at zero. The LSTM ran for 235 epochs before stopping due to a lack of further improvement in the validation set error. The MSE's were 0.019 for the ARIMA prediction and for 0.002 for the LSTM.

\begin{figure}[!ht]
\centering
\includegraphics[width=.8\columnwidth, height=1.7in]{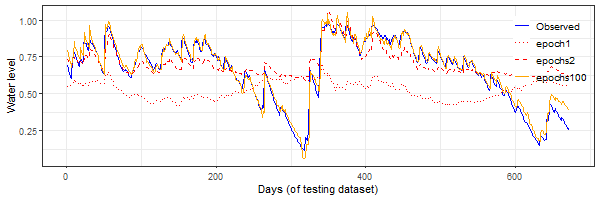}
\caption[short title]{Evolution of predicted groundwater level on testing data set after 1, 2, and 100 training epochs of a single-layer LSTM, with 32 nodes, 50 lags, and 1e-6 weight decay. (Note that this is a different generated data set to the one above.)}
\label{fig:trainEvolve}
\end{figure}

During training of the LSTM network, the evolution of the prediction can be traced through the epochs. Figure \ref{fig:trainEvolve} shows the estimated groundwater level after 1, 2 and 100 epochs of training the network. We can see that after only a couple epochs the predicted time series was approaching the upper values of the observed data, but the lower values were not yet being captured. By 100 epochs, the upper and lower values were both being captured well.

\section{Application} 
\label{application}
We now turn to an application that involves the modelling of water levels in a monitoring bore located in the Richmond River catchment in eastern Australia. The region has a sub-tropical climate with a clear wet/dry seasonal pattern, though not
as extreme as the rainfall patterns from Tully, used for the simulation section of the paper.  We expect
that the major challenge in developing a reliable groundwater prediction model will be in constructing a predictor space that is rich enough to capture not only potentially long response time lags, but also non-linearities in the relationship between the borewater levels, rain and other climate variables. We will use ARIMA and LSTM models to predict groundwater levels for known rainfall and evapotranspiration sequences, and the predictions will be compared with the corresponding measured water levels.

\subsection{Data} 

Bore water level data were obtained from a publicly accessible database that contains information on rivers, various ground water sources, dams and aquifers throughout the state of NSW (see \url{https://realtimedata.waternsw.com.au/}).  Information provided through this site is gathered from a combination of private and public sources and ranges from periodic measurements reported by private citizens to official monitoring bores that are equipped with telemeters that collect data on an almost continuous basis. Groundwater level data was obtained from this database for site GW041001 (borehole 1, pipe 1) in the Alstonville region. Recorded hourly values were converted to daily means for use in the models.

 Climate data measured at the Australian Bureau of Meteorology station 58131 (also located in Alstonville) were obtained from the SILO database ((ref: https://www. longpaddock.qld.gov.au/silo/, station 32042)). A continuous time series of data with no missing values are available at this station for 14 years (2006-2020).  

\begin{figure}[!ht]
    \centering
 \includegraphics[width=4in]{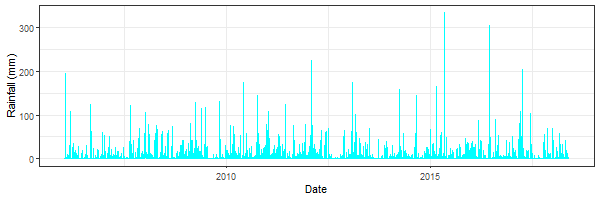}\\
 \includegraphics[width=4in]{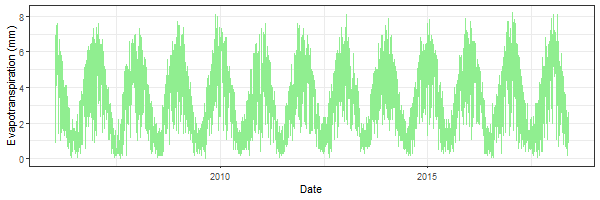}\\
 \includegraphics[width=4in]{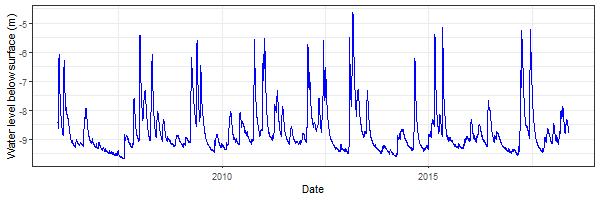}
    \caption{Observed data at Alstonville measuring station.}
    \label{fig:obs_Alston}
\end{figure}

The climate and borehole data used in this application are shown in Figure \ref{fig:obs_Alston}.

\subsection{Modelling}

Following on from methods described in Section 3, we ran LSTM models  with a series of hyperparameter combinations to determine an appropriate blend of the number of lags, number of nodes, number of layers, weight decay and patience. A total of 2880 model runs were performed with various hyperparameter permutations.  Runs were performed on a high performance computing cluster. 

\begin{figure}
\subfloat[]{\includegraphics[width = 3in]{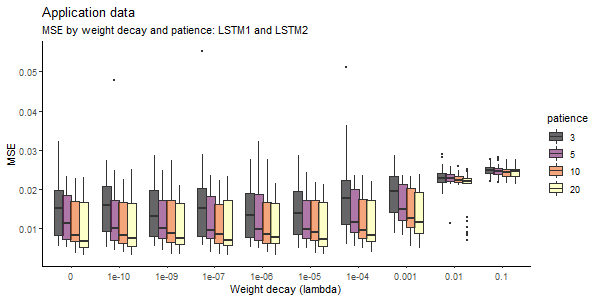}} 
\subfloat[]{\includegraphics[width = 3in]{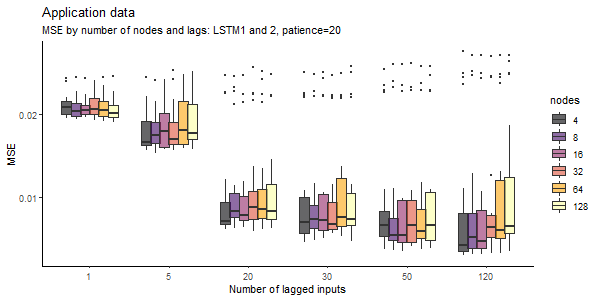}}\\
\subfloat[]{\includegraphics[width = 3in]{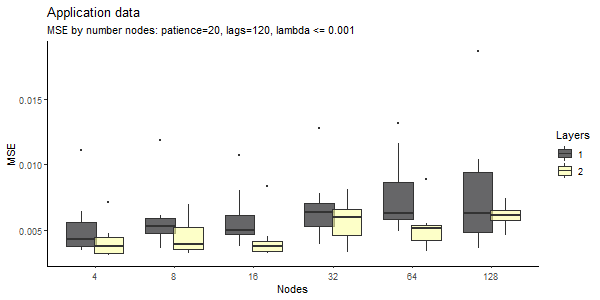}}
\subfloat[]{\includegraphics[width = 3in]{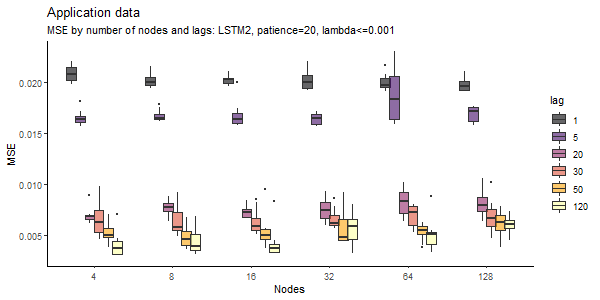}} 
\caption{Mean squared errors with various hyperparameters for application LSTM.}
\label{fig:appl_hyp}
\end{figure}

The boxplots in Figure \ref{fig:appl_hyp} show the MSE's for models run with different hyperparameter combinations. From (a) we see that the weight decay parameter has an influence on MSE, with values of lambda less than 0.001 producing the best models. We also see from this plot that the patience value of 20 consistently produces the lowest MSE's when compared with the other patience values. In plot (b) we look at results of models run with patience of 20, for groupings of lags and nodes. It can be seen that the models including 50 or 120 lagged values have the lowest errors, and that errors do not generally decrease with increasing nodes. It appears that 50 lagged values with less than 32 nodes, or 120 lagged values with less than 16 nodes may provide the best combinations. Plot (c) breaks the series of runs up by number of nodes and by number of LSTM layers. Here it is clear that 2-layer LSTM's are producing the best results. Having narrowed the best runs down to 2-layer networks with regularisation parameter lambda less than 0.001, run with patience of 20, we now see from the results for this subset of runs in plot (d) that 16 nodes and 120 lagged values will provide a good model setup.

Using this LSTM setup, we predict the groundwater level as a function of rainfall and evapotranspiration for the testing portion of the Alstonville time series. The same is done with an ARIMA model, also including the previous 120 days of rainfall and evapotranspiration as lagged predictor values. 

\subsection{Aquifer level prediction results}

\begin{figure}[!ht]
\centering
\includegraphics[width=.8\columnwidth]{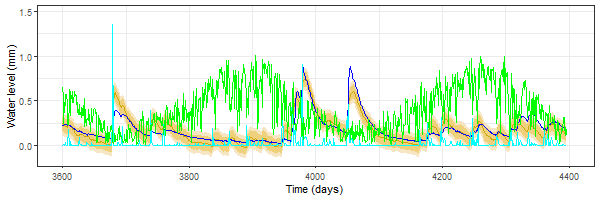}
\includegraphics[width=.8\columnwidth]{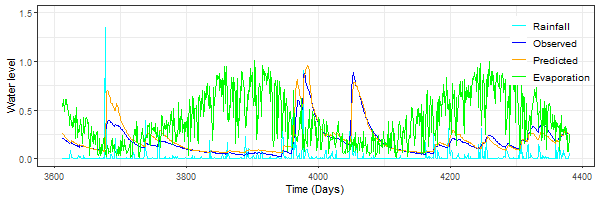}
\caption[short title]{Predicted groundwater level response with ARIMA(2,0,2) (upper plot) and LSTM (lower plot) models. The ARIMA prediction includes 80\% and 95\% prediction intervals. Both models include the 120 preceding days of rainfall and evapotranspiration measurements as lagged predictor variables. The mean squared errors for both predictions, compared with the observed water level timeseries, are 0.004.}
\label{fig:Alstonboth}
\end{figure}

Groundwater level timeseries predictions are produced using the ARIMA and the LSTM models based on rainfall and evapotranspiration in a testing portion of the measured timeseries. The results are shown in Figure \ref{fig:Alstonboth}, and are compared to the observed groundwater level time series in this portion of the data set. One thing we notice about this particular portion of the data is that there is a spike in rainfall near the beginning which reaches to (a scaled value of) 1.35. This is well beyond the range [0,1] in which the scaled training data for the LSTM resides, however, the training data set for the ARIMA includes rainfall of 1.48 as there is no data reserved for validation when using the ARIMA model and the first 80\% of data comprises the training set (compared with 60\% for the LSTM). The LSTM model has therefore not been trained on rainfall data this large and understandably struggles to replicate the observed conditions at this point.

The prediction produced with the ARIMA model includes 80\% and 95\% prediction intervals for the predicted water level. The best ARIMA model was found to be ARIMA(2,0,2) based on AIC in the training data set, with $d$ set to zero. The prediction has a mean squared error of 0.004 in the testing data set. The 2-layer LSTM model with 16 nodes on each layer, patience of 20, and lambda of $10^{-5}$, ran for 94 epochs and produced a prediction also with an MSE of 0.004 in the testing data set.

\begin{figure}[!ht]
    \centering
\includegraphics[width=.8\columnwidth, height=1.7in]{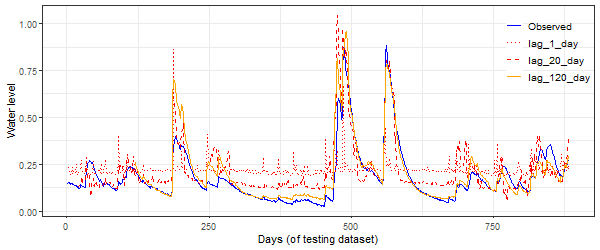}
    \caption{Influence of number of lags on prediction accuracy in the LSTM: predicted water levels are shown for models run with different numbers of lagged days included in the predictor variables.}
    \label{fig:AlstonLags}
  \end{figure}

 The number of lagged days of predictor values included in the model has a great effect on the prediction accuracy, as we have seen in Figures \ref{fig:LSTM1_hyp} and \ref{fig:appl_hyp}. Table \ref{MSEtable} lists the MSE's resulting from runs of the ARIMA and LSTM models with different numbers of lagged predictor values included. The 'best' ARIMA model structure ($p,d,q$) and the number of epochs the LSTM runs through before stopping due to lack of further improvement are listed. The ARIMA models are restricted to those with $d=0$. Except for the LSTM with 1 lag, LSTM models including higher numbers of lagged inputs ran for fewer epochs before converging. In Figure \ref{fig:AlstonLags} we visualise how varying the quantity of lagged predictor values influences the results. Shorter lags lead to predicted time series that are flattened compared with the observed timeseries, and therefore have higher MSE's. Peaks and troughs of the observed data are captured best when longer lags are included. 

\begin{table}[!ht]
 \caption{Results of a series of model runs with varying lags \ \label{MSEtable}}
\centering
\begin{tabular}[b]{c | cc | cc}\hline
 \hline
  & ARIMA & ARIMA & LSTM & LSTM\\
 Lags & Best model & MSE & MSE & Epochs\\ 
 \hline
1 & (2,0,2) & 0.021 & 0.027 & 20 \\ 
5 & (2,0,4) & 0.019 &  0.016 & 398\\
20 & (2,0,2) & 0.009 &  0.008 & 305\\
50 & (3,0,3) & 0.005 &  0.007 & 145\\
120 & (2,0,2) & 0.004 &  0.004 & 94 \\
\hline
    \end{tabular}
    \end{table}

\begin{figure}[!ht]
    \centering
\includegraphics[width=.5\columnwidth]{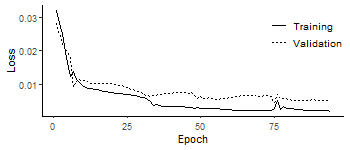}
    \caption{Mean squared error for training and validation data sets over the 94 epochs of LSTM model training on the real world data set. }
    \label{fig:ApplLoss}
  \end{figure}

 Mean squared error losses are recorded after each epoch of LSTM model training, both for the training data set and the validation data set. The results for the 2-layer LSTM used in this application are shown in Figure \ref{fig:ApplLoss}. Training was stopped when the MSE for the validation set failed to improve over 20 epochs. It can be seen that if the patience had been set to a lower value, training may have halted at around 10 epochs where there was a dip in the validation loss.
 
\section{Discussion}
\label{discussion}
In this paper, we use an application in groundwater modelling as the basis for 
discussing and comparing classical statistical approaches to time series as well as more recent developments in neural network modelling, or deep learning, from the machine learning literature. 
From a high-level perspective, our problem can be described in fairly simple terms as
predicting one time series $Y_t$ (in our case, bore water levels) as a function of a
vector of other relevant time series $\bX_t$ (in our case, rain and evapotranspiration).  

But as one often finds with real world applications,``the devil is in the details" and things get complicated quickly.  Groundwater aquifers are recharged by rainfall in a complex manner depending on numerous factors, including the geological structures and nature of the rocks and soil below the surface, as well as other climatic factors such as temperature and humidity.  If an aquifer is shallow and the soil above it fairly porous, then rainfall may recharge it fairly quickly, in a matter of days.  In contrast, it might take weeks or even months for rainfall to reach deep aquifers or even shallow ones encapsulated by rocks or less permeable soils.  

There is an extensive hydrological literature focused on how to build reliable models that capture all of these complex relationships.   Many rely on open source software called {\tt modflow} developed by the US Geological Survey \citep{modflow} that is based on hydro-dynamical theory and relies on specification of the shape and extent of the aquifer and the nature of the soil and rocks surrounding it. Collecting the data needed to build a reliable model will generally be expensive and time consuming. \cite{BakkerSchaars} refer to these physics-based models as ``white box" since all
the assumptions and inputs are very clearly laid out.  In contrast, they refer to the machine learning
as ``black box" since these approaches give answers without offering much direct insight into
the relationships between the predictors and outcomes.  
\cite{BakkerSchaars} use the term ``grey box" to refer to more classical statistical approaches
which try to engage with subject matter experts to get a sense of what's important and use that information to build up a suitable set of predictors, whilst still letting the data help determine
the precise nature of the relationships. 

The use of neural networks in water resources has been reviewed in detail by \cite{shen2018transdisciplinary}. Applications include dynamic modelling of sensor data, learning of data distributions, analysis of raw images from remote sensing, and the potential identification of unrecognized or unrepresented linkages between system components. The author notes how few studies have been done using deep learning in hydrological applications, and this review paper attempted to cover every application done to date in this field.  

The LSTM has proved successful in the hydrological applications in which it has been used, especially those in which data is plentiful but information on the physical processes is limited. \cite{zhang2018developing} found the LSTM to be a valuable tool for water table depth prediction in regions where it was difficult to obtain hydrogeological data, or the known hydrogeological characteristics were particularly complex. \cite{lee2020stochastic} found the long-term variability and correlation structure of streamflow systems well preserved by the LSTM over annual timescales. \cite{kratzert2018rainfall} demonstrated that the LSTM was better at determining long-term dependencies (runoff in spring based on snowfall in winter) than the RNN. They were able to predict streamflow at hundreds of basins without the need for including any differentiating physical characteristics.  \cite{jeong2019comparative} used an LSTM built on data from a certain time period to determine if the system was experiencing stress at a later time.

A very important attribute of neural networks with respect to hydrology, and other physical sciences that rely heavily on automated sensors for data collection, is their inherent ability to perform feature extraction on the raw data. This enables the addition of new data to the model in real time as it emerges from sensors, without any requirement for transformation of the entire data set or manual feature construction. Once the model is built, predictions can be updated in real time as new data comes in.

Differences between classical and machine learning approaches have long been pointed out. Neural networks are more concerned with accurately modelling systems than providing insight into why they behave the way they do, as is done with classical approaches. There are benefits and drawbacks of each approach, in terms of both required input effort and output results, which have been seen in this study.

Neural network approaches are very appealing in terms of the potential to capture non-linearities and interactions without requiring an explicit mathematical understanding of the underlying system. Neural networks also have the advantage of engineering the predictor space. However, the implementation is non-trivial. Any single run is straightforward, but many runs are required to tune and select the appropriate setup parameters. Neural networks are stochastic models due to the random initialisation and optimisation process, and therefore can produce varying results with every run on the same data. Because of this, even after the setup parameters are tuned, one will usually end up training numerous models on the same data and choosing the one with the best prediction.

On the other hand, the ARIMA approach is more principled in terms of model fitting. There are no corresponding setup parameters such as 'early stopping' or number of layers to specify, and the enhancement of the prediction with confidence intervals is of great value.  But, the challenge with ARIMA modelling is to make sure that the assumed predictor space is rich enough to capture the important relationships needed to provide good predictions. We 
saw in our simulations that the ARIMA models, and also the linear neural network models, did not do as well 
as the more general neural network models that had the flexibility to incorporate non-linear relationships. It is important to note, however, that ARIMA models can encounter computational instability in settings where the specified predictor space is large.  In our application, for example, we took the relatively naive approach of including 120 lagged rainfall  variables in our model.  As it turned out, the model seemed to fit well.  However, it is easy to imagine that this approach might lead to multi-collinearities.  There are obvious strategies that could be used to help, including penalization or through the use of distributed lag models \citep{Koyck}.  As discussed in the methods section, linearity assumptions can be relaxed for the ARIMA model through the inclusion of additional terms such as regression splines.  However, computational instabilities could easily arise in settings involving multiple lagged predictors.  \citep{wood} discusses the use of bivariate smoothing methods to fit flexible distributed lag models, though this
only works in the setting of independent errors.   We believe that this could be a useful avenue for further exploration.

 Ideally, one would like a hybrid of the sound statistical framework for inference, but combined with the power of feature engineering provided by the neural network approach. In practice, analysts sometimes extract the top layer from a fitted neural network model and use those elements as input to more statistical approaches.
  All this being said, it is very surprising that the relatively simple ARIMA approach did so well on our real world application. In future, it would be worthwhile expanding our exploration to other monitoring bores which perhaps exhibit more complex patterns over time. 
 Also, the natural benefit of recurrent neural networks would be better realised when exploring relationships between many more related time series than we have used here, as the LSTM allows non-linear dependencies among multiple time series to be found. We could for instance consider exploring the possibilities of including more climatic variables, adding a spatial aspect by predicting water levels from data collected at neighbouring bores, and incorporating anthropogenic influences on water levels such as extractions
and dam releases.

\section*{Acknowledgements}
Clark, Hyndman and Ryan are all supported by the Australian Research Council 
through the Australian Centre of Excellence in Mathematical and Statistical 
Frontiers (ACEMS; CE140100049).  Pagendam is supported by CSIRO which is an industry partner of ACEMS.

\bibliographystyle{chicago}
\interlinepenalty=10000
\bibliography{biblio}

% \newpage
% \begin{table}
%  \caption{Results of a series of model runs with varying lags \ \label{MSEtable}}
% \centering
% \begin{tabular}[b]{c | cc | cc}\hline
%  \hline
%   & ARIMA & ARIMA & LSTM & LSTM\\
%  Lags & Best model & MSE & MSE & Epochs\\ 
%  \hline
% 1 & (2,0,2) & 0.021 & 0.027 & 20 \\ 
% 5 & (2,0,4) & 0.019 &  0.016 & 398\\
% 20 & (2,0,2) & 0.009 &  0.008 & 305\\
% 50 & (3,0,3) & 0.005 &  0.007 & 145\\
% 120 & (2,0,2) & 0.004 &  0.004 & 94 \\
% \hline
%     \end{tabular}
%     \end{table}

% \newpage
% \input{figures.tex} 
\end{document}